\documentclass[numbers=noendperiod,chapterprefix,sopenany,fontsize=11,oneside,bibliography=totoc,parskip=half-]{scrartcl}
\usepackage{epsf, pstricks, graphics, graphicx} 
\usepackage[a4paper]{geometry}
\usepackage{multirow} 
\usepackage{tikz}
\usepackage{dot2texi}
\usepackage{svg}
\usepackage{physics}
\usepackage{endnotes} 
\usepackage[edges]{forest}
\usepackage[colorlinks=true, urlcolor=blue, linkcolor=blue]{hyperref} 
\hypersetup{colorlinks = true, urlcolor = blue}
\usepackage{xfrac} 
\usepackage[backend=biber, citestyle=numeric, bibstyle=numeric,sortcites]{biblatex}
\usepackage{float} 
\usepackage{pbox} 
\usepackage{hyperref} 
\usepackage{threeparttable}
\usepackage{rotating} 
\usepackage{setspace} 
\usepackage{rotating}
\usepackage{multicol} 
\usepackage{tablefootnote}
\usepackage{cleveref}
\usepackage[T1]{fontenc} 
\usepackage[utf8]{inputenc}
\usepackage{amsmath}
\usepackage{mathtools}
\usepackage{amssymb}
\usepackage{xfrac}
\usepackage{mathcomp}
\usepackage{lscape}
\usepackage{imakeidx}
\usepackage{tabularx}
\usepackage{multicol}
\usepackage{framed}
\usepackage[german,english,polutonikogreek]{babel}
\newcommand{\x}[1]{\textit{#1}}

\newcommand{\mb}[1]{\mathbf{#1}}

\begin{document} 
\selectlanguage{english}
\title{Ontologies of common sense, physics and mathematics}
\author{Jobst Landgrebe \and Barry Smith}
\date{\today}
\publishers{University at Buffalo\\
jobstlan@buffalo.edu, phismith@buffalo.edu}
\maketitle

\begin{abstract}
The view of nature we adopt in the natural attitude is determined by common sense, without which we could not survive. Classical physics is modelled on this common-sense view of nature, and uses mathematics to formalise our natural understanding of the causes and effects we observe in time and space when we select subsystems of nature for modelling. But in modern physics, we do not go beyond the realm of common sense by augmenting our knowledge of what is going on in nature. Rather, we have measurements that we do not understand, so we know nothing about the ontology of what we measure. We help ourselves by using entities from mathematics, which we fully understand ontologically. But we have no ontology of the reality of modern physics; we have only what we can assert mathematically. In this paper, we describe the ontology of classical and modern physics against this background and show how it relates to  the ontology of common sense and of mathematics.
\end{abstract}





Before Kepler and Galileo, physics was a science based on arithmetic and trigonometry and with a correspondingly limited reach. Today, however, physics provides the theoretical foundation of the technosphere in which we live and it provides our models for the understanding of the universe and matter.
Experience teaches us that we can use the laws of physics to create reliable technological artefacts such as cars, planes and MRI devices. Such devices have enhanced human lives to a degree which would have been unimaginable in former times.
How did this evolution take place? And why is the physics we have today so useful, given that, as we shall see, the models it uses are all in a strict sense wrong?\footnote{We here follow the arguments of \textcite{cartwright:1983} to the effect that the models of physics are either merely approximations of reality made under certain simplifying conditions or such as to apply only to certain restricted systems which are either deliberately chosen from the vast realm of inanimate nature or built artificially. A model is a mind-dependent representation of an aspect of reality using  abstract symbols (and potentially also text and figures) that is created to describe, explain, or predict the aspect of reality in question. More details are given in section \ref{models}.}
The attempt to answer this and related questions has led to a huge body of work in the philosophy of science and quantum physics.\footnote{For examples see: \cite{butterear:2007, esfeld:2012, esfeld:2017, lewis:2016}.}
Missing from all of this work, however, is any systematic attempt to develop an \x{ontology} of physics.

An ontology in the sense used in this paper is a formal representation of  `the kinds and structures of objects, properties, events, processes, and relations' in some domain of reality.
The applied ontologist seeks to provide a definitive and exhaustive classification of entities in specific domains and of the relations between them \cite{smith:2003}.
Such classifications have become increasingly important in many areas. They are now used routinely in the life sciences \cite{smith:2023, gene:2021}, where they promote reuse and exchange of data deriving from different, heterogeneous sources by providing common systems for data annotation.

Why, then, has no convincing ontology of physics been created thus far?
\x{First}, physicists and applied mathematicians working in physics do not feel the need for such an ontology. 
This is because they already have \x{mathematical} models of reality which have been validated over and over again for the established body of knowledge in physics through experiments.
These models are well understood and accepted as a matter of course by all major groups of physicists and the associated experimental data are annotated using commonly accepted mathematical formalisms. This means that there is no major impediment to the communication between different communities of both theories and data -- thus no impediment of the sort that we find in the life sciences and in other areas where work in ontology plays an important role.
Nowadays, indeed, the experiments in physics are themselves often planned and conducted by huge international consortia such as CERN, ISS or ITER, with the aim of ensuring consistent collection and unproblematic exchange of data across national and linguistic boundaries.

\x{Second}, physics is distinguished from biology or chemistry by the fact that, where the latter are descriptive sciences dealing with billions of different types of entities, physics is a systematic science describing general principles\footnote{A \x{principle of physics} is a highly abstract, universal type of relation between material objects that is represented in corresponding models.
Examples are the principle of least action or the symmetries represented in models of nature by symmetry operations such as translations in time or space.} and laws of nature.\footnote{A law of nature is a mathematical model of an interaction of material system elements that has universal validity. Examples are the models defined by the field equations of the General Theory of Relativity or by the Schr{\"o}dinger equation for quantum systems with a small number of particles.}

And \x{lastly}, relationships found in one domain of physics are frequently reused also in other domains, for example where analogues of the phenomenon of mechanical resonance (forced oscillator with dampening) reappear in the phenomenon of electrical resonance.
Chemistry can be seen as a branch of physics, and physical properties\footnote{Here in the sense of: matter-related properties, by which we mean, when not stated otherwise, physics-(science)-related. \cite[sect.~2.2.2]{landgrebesmith:2022}} of the elements have been used as basis for classification systems of chemical entities such as PubChem or ChEBi \cite{hastings:2012}. Furthermore, classification in physics is often performed using mathematical structures. For example, in quantum field theory, the mathematics of group theory has been used to create the classification scheme that now forms the basis of the standard model of particle physics \cite{weinberg:1995}. 
For these reasons, the classification problems of physics are, already to a large degree solved. 

The need for an ontology of physics arises, rather, at the point where physics is \x{applied}, for example, in materials science, which deals with millions of different types of entities which need to be taxonomically organised in order to enable data exchange and system interoperability in areas such as solid state engineering. 

However, to create physics domain ontologies which address this need in a way that can promote interoperability\footnote{This is the ability to exchange data between machines and enable automated computation (search, aggregation modelling) on the data.}, a common upper-level ontology of physics is required which defines the basic kinds of entities and relations that applied scientists and engineers have to deal with\footnote{For example, see https://industrialontologies.org/iof-charter/}. 
We proceed as follows. First, we briefly state our view of the nature of theories in physics and describe our proposed upper ontologies of physics (PhysO) and mathematics (MathO) and document also how they relate to the BFO ontology of common sense.\footnote{This is a top-level ontology used especially in the life sciences which has been documented as an international standard in ISO/IEC 21838-2. See \cite{arp:2015}. The authors of BFO did not see the need to include mathematical entities in the coverage domain of the ontology. In particular the authors of BFO did not include any room in the ontology for the treatment of sets.}
We then sketch applications of these upper ontologies to two highly contrasting examples, one from the domain of classical mechanics (the harmonic oscillator), the other from quantum physics (photon entanglement). Our rationale for this choice of examples is that it will allow us to demonstrate that phenomena from two of the most important domains of physics, domains which require fundamentally different mathematical formalisms, can be dealt with coherently within a single ontological framework. Success in this regard will, we believe, lend support to our hypothesis that all other domains of physics can be described using the proposed framework.
Based on these examples, we then summarise our view of the relationships between the ontologies of, respectively, common sense (BFO), physics (PhysO) and mathematics (MathO).

\section{Ontology of physics}
\subsection{The nature of theories and models in physics}

We view physics as the science of inanimate natural and technical systems, where a system, in the sense relevant to our purposes here, is a totality of dynamically interrelated material elements. 

Systems are parts of nature delimited by fiat.
To delimit a system is to select a level of granularity of its elements, from microphysical particles to entire galaxies, and of the interactions between these elements, from gravitational attraction to galactic collision. It is typically also to select a boundary around those elements which will form the system in question. Physics takes such delimited systems as its object because the complexity of nature in its undelimited totality would go far beyond what can be achieved by any theoretical approach. 

\subsubsection{Classical and modern physics}
Classical physics is the physics that was pursued from the inception of the discipline in the early 17th century (Kepler, Galileo) to 1905 -- the decisive year in which Planck described the quantization of energy levels of the electron and Einstein the special theory of relativity. These two discoveries  mark the beginning of modern physics, which is distinguished from classical physics through its use of abstract mathematical structures which cannot be imagined using the natural attitude: Hilbert spaces for quantum mechanics and Riemanian manifolds for the general theory of relativity.

\subsubsection{Laws as models}\label{modelsLaws}
Since the late 18th century, it is mathematical equations which generate the descriptive, explanatory or predictive power of the laws of physics. 
Equations can express two sorts of laws: \x{fundamental} and \x{phenomenological} \cite[essay~6]{cartwright:1983}. 
A \x{fundamental law} is a conjunction of one or more equations created to contribute to a model that (i) describes the behaviour of systems of a certain type and (ii) is `universal' in the sense that it describes the behavior of a system that is found throughout the universe and thus also models something that is independent of the human mind.
Examples are the equations of Einstein's field theory or Maxwell's equations of electromagnetism. Although such equations are sometimes also called `causal laws', they do not in fact \x{cause} anything. Rather, they \x{describe} causal relationships. Both principles and laws are, in our view, mere models. The first have explanatory power which involves an appeal to causality; the second have predictive power.

A \x{phenomenological law} is a model of inanimate nature\footnote{Nature is the totality of material entities of the world not created by humans; the totality of what is created by humans forms culture (\textgreek{t'eqnh| >'ontes} in the sense of Aristotle).} describing the behaviour of some specific type of natural system, and are thus without universal validity. 
An example is the conjunction of equations expressing the quantum theory of the laser. 

Models in physics typically contain not only equations and figures but also text defining the assumptions of the model, for example in descriptions of a natural system using `as if' clauses  (as in: `helium gas behaves \x{as if} it is a collection of molecules which interact only on collision' \cite[p.~128f.]{cartwright:1983}). 
Text of the sort that occurs in models may be used also to justify the fundamental laws of a theory, or to justify the use of mathematical approximations introduced to make the equations of the model solvable (for example, by neglecting terms of equations under conditions specified in the text).

\subsubsection{The scope of physics}
Each of these cases illustrates how the most prestigious models in physics \x{idealize reality}. Indeed, the fundamental laws of physics are not, as was supposed by 17th century authors such as Boyle or Hooke, laws written by God into the book of nature. Rather, as Cartwright puts it, they model simplifying abstractions of natural processes; they `do not govern objects in reality; they govern only objects in models.' \cite[p.~129f.]{cartwright:1983}  In this connection Cartwright speaks of `physics as theatre' (p.~139).
The laws are indeed themselves mathematical models of nature. But they do not accurately describe the complexity of nature itself; nor can we learn from them what nature is really like. Hence verbal descriptions in physics often draw attention to the fact that a representation of the full complexity of the systems under study is impossible, followed by documentation of the simplifying assumptions.

But we do of course learn \x{something} from models: What they provide is a consistent and, with regard to the mathematical part, logically structured approximation of certain aspects of nature, above all of those which allow us to make predictions about systems and events that are relevant to the realization of human goals because they allow the creation of technical artifacts. In other words, they allow us to explain\footnote{An explanation is the description of the causality of a system. For example, Newton's laws explain the movements of the planets in the solar system by describing them causally.} aspects of nature in ways that are instrumental uses of physics.\footnote{Our view is thus in contrast to Cartwright's later view that fundamental laws do not apply to nature at all \cite[ch.~1]{cartwright:1999}.}

\subsubsection{The need for approximations}
We must use approximations, in all of this, because there is a discrepancy between the causality of nature and our ability to do justice to this causality when we formalize the laws of nature using mathematics. This arises because the causalities are too complex to model exactly using mathematics.
This is in conformity with our assumption, widely accepted among physicists, that all events in nature are caused by combinations of the four fundamental interactions of electromagnetism, gravitation, and the strong and weak forces.\footnote{Where the first two are the only forces we can observe at the mesoscopic level of granularity of common sense, the views of physics on the weak and strong forces are extremely well grounded in experiments and theory, though our assumptions concerning the fundamental role of the four forces may change in the future if different paradigms for experimentally analysing matter would become available.} Sound, for example, can be reduced to the electromagnetic interactions of the particles involved in a sound wave travelling through a medium. 
For some natural systems, the behavior can be formalized by means of universal laws using these interactions to describe the relationships between their elements. The assumption of fundamental laws of physics is powerful for prediction and explanation especially for those natural systems whose behavior can be observed unproblematically, as was shown most consequentially by Newton in his treatment of the solar system as a gravitational system. This assumption can yield valuable results also for those natural systems in relation to which we can perform laboratory experiments. However, for the vast majority of natural systems, including nearly all systems in organic nature, we cannot model mathematically the ways this causation occurs \cite[chapter~8]{landgrebesmith:2022}.
Thus while the common-sense view of nature, central elements of which are reflected in Basic Formal Ontology, is very powerful -- it provides us with all of the knowledge that is relevant to our daily lives -- it does not extend into the realm of physical science.

\section{Physics upper ontology}

The PhysO ontology we propose can therefore not be an ontology of nature \x{simpliciter}. Rather, it is an ontology of nature \x{as it is modelled mathematically} at different levels of abstraction -- such as classical geometry and calculus for classical physics, and abstract mathematical entities out of scope for common sense for modern physics. The ontology of physics thus depends always on the ontology of mathematics. 

\begin{figure}[htb]
\begin{minipage}{\linewidth}
\begin{footnotesize}
\begin{tabular}{c}
\begin{forest}
forked edges,
for tree={draw,align=center}
[physics\\ entity
  [system\\ entity
   [system\\ element
    [weight]]
   [system\\ relation
    [electro-magnetic\\ interaction]]
   [system
   [weight-on-\\spring]]]
   [magnitude
    [continuant\\ quality
     [mass]]
    [process\\ characteristic
     [acceleration]]]
   [model
    [textual\\ model]
    [graphic\\ model
      [geometric\\ model
	     [orbit\\ model]]]
    [mathematical\\ model 
     [forced oscillator\\with dampening]]]
    ]
\end{forest}\\
\end{tabular}
\caption{Top-level entities of PhysO, lowest nodes here show only examples, which are documented in section \ref{ho}\label{f:ontP}. Note that the system entities and the magnitudes shown here can also be classified using BFO, and that the mathematical models consist of representations of mathematical entities classified using the mathematics ontology described in section \ref{matho}.}
\end{footnotesize}
\end{minipage}
\end{figure}
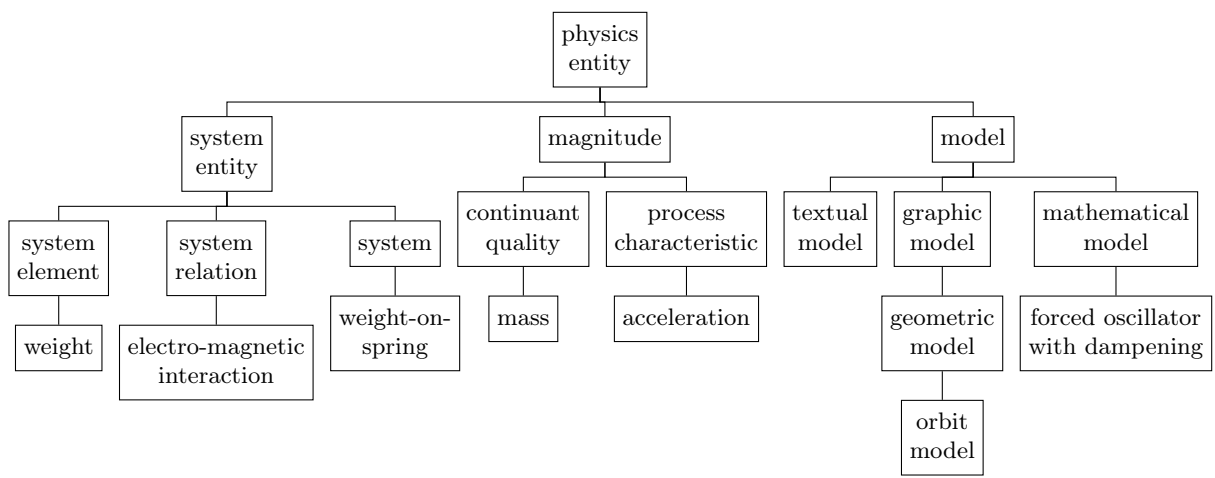

The ontology of physics comprises three main axes: system entities, magnitudes, and models.
An overview is shown in figure \ref{f:ontP}, where we follow the standard strategy of representing an ontology as a directed acyclic graph with nodes representing entities and edges representing binary relations between these entities (in this figure: Aristotelian genus-species (class-subclass) relations\footnote{Often desginated as `is\_a' relations in the applied ontology literature.}).

Importantly, in classical physics, the system entity branch of the physics ontology falls within the domain of the common-sense ontology BFO. In modern physics, in contrast, no branch contains BFO entities.\footnote{See section \ref{ontorel} and figure \ref{f:venn} for more details.}

We now proceed to describe the three branches of the physics ontology.

\subsection{System elements} \label{systems}

A system is a part of reality. It is a totality of entities (called `system elements' in figure \ref{f:ontP}), each made of matter, which interact with each other; thus of elements which participate in processes of interaction. To delimit a system is to select a level of granularity of elements and a system boundary. For example, the solar system can be seen as a gravitational system in which the elements interact via the force of gravity. Its elements are planets, satellites and other objects bound by the sun's gravitational force. 

Different systems are delimited according to what are taken as elements: just the planets; or the planets together with their respective satellites; or also pieces of space debris bound by the sun's gravitational force.
As these examples again make clear, systems are always delimited by fiat. The sun itself can also be seen as a system of electromagnetic interactions, whose elements are the particles emitted by the sun.

System elements can range from elementary particles to entire galaxies. The interactions between the elements are governed in every case by one or more of the four fundamental forces listed above (see `system relation' in figure \ref{f:ontP}).\footnote{We have inductive, positive knowledge about the four forces, but we know little about their fundamental nature. We observe them, we understand and model them as causes of events, but in the end, as Feynman puts it, we do not really know what a force is \cite[ch.~12]{feynman:2010}.}
We can observe and measure each of the four forces and physicists have not identified any other force that is not composed of them. When several effects of one force or several forces are overlayed in complex systems, it is in general impossible to contrive mathematical models that are of sufficient quality for exact description or high-quality prediction \cite{landgrebesmith:2022, thurner:2018}. Examples we have from inanimate nature, such as the solar system as a gravitational system, are very close to what we have called logic systems \cite[chapter~7]{landgrebesmith:2022} and can be modelled very well using mathematics.

\subsection{Magnitudes} \label{magnitudes}

In classical physics, a \textit{magnitude} is a phenomenon in reality -- for example mass, distance, velocity, acceleration or energy -- which has the feature that its instances can be \x{measured}. The phenomenon in question exists both on the level of instances and on the level of universals.\footnote{A universal is an abstract entity that is instantiated in reality by an indefinite number of instances. We gain knowledge of universals by abstracting from their instances, for example, going from  this concrete atom of hydrogen in this water molecule in this glass to the universal \x{`hydrogen atom'}. \label{hydrog}}

Classical physics magnitudes, the associated processes of measurement, and the underlying qualities are all such as to fall within the realm of what can be imagined by common sense (that is, in the realm of Basic Formal Ontology). The magnitude universals of non-relativistic mass, time, distance, and so forth, have existed since the beginning of the universe, and thus long before any measurements occurred. The \x{mathematical} treatment of magnitudes, on the other hand, is mind-dependent. For example, the terms `$F$', '$m$' and `$a$' in the equation `$F=ma$' which are all common-sense magnitudes in classical physics, when used in a mathematical context are mere variables, \footnote{A variable (as contrasted with an associated symbol) is a mathematical entity which serves as typed placeholder in a mathematical term ranging over a set of mathematical entities. `Typed', here, means that the typed placeholder can only take values from its associated set.\label{variab}} and they are treated as such for example when you use `$F/m$' to substitute for `$a$' in another equation. 

In modern physics, in contrast, for example in quantum physics and the general theory of relativity (GTR), magnitudes no longer fall within the domain of what can be imagined using common sense. In GTR, for example, \x{time} is conceived as one inseparable dimension of a four-dimensional spacetime manifold, a structure adapted not from our experience of reality -- either in common sense or in experiment -- but rather from mathematics. Thus time in GTR \x{is not a universal} in the sense outlined in footnote \ref{hydrog}, and it is not something that we measure like we measure distance in classical physics.
We can certainly identify \x{particular times} in GTR, such as the moment a supernova begins to form. But we cannot separate it from the spacetime manifold and consider it as something existing in indefinitely many interchangeable copies. Moments in time in GTR go hand in hand not with a framework of universals of the sort with which we are familiar in our common-sense experience, but rather with a structure that is mathematical in nature -- a structure that we ourselves make.

In GTR \x{the time of classical physics, the time compatible with common sense, is gone.} Similarly, the quantum-theoretical magnitude \x{spin}\footnote{See section \ref{m:phot} below.} is a mathematical entity used to model measurement results of magnitudes we cannot understand using our common-sense-based view of the world or our natural (non-mathematical) imagination.

As for time in GTR, so also for spin, we have measurements (spin direction) for particular particles, but no corresponding universal. We thus have the following situation:

\begin{itemize}
 \item classical physics: $\langle \mathrm{universal-instance} \rangle,$
 \item modern physics:  $\langle \mathrm{mathematical~entity-measurement} \rangle$,
\end{itemize}

where the instance of classical physics is replaced in modern physics by observation or measurement. The second type of tuple is a \x{amatter of artificial declaration} (attribution)\label{declared}.\footnote{A \x{tuple} $\langle x,y\rangle$ is an ordered pair, in set-theory it is written as follows: $\{\{\varnothing, \{x\}\}, \{y\}\}$.}
This is because, when we measure, for example, spin, there is nothing which we can identify as instance of a universal. In other words, the tuples of type $\langle \mathrm{mathematical~entity-measurement} \rangle$ are created by declaration because we are unable to understand the phenomena related to their associated magnitudes using common sense, we cannot think of them as universals by abstracting from their instances as we do, for example, when we start out from individual beeches and oaks to obtain the universal `tree'. 

Therefore, modern physics has declared that a mathematical entity (for example a structure like the manifold in terms of which spacetime is defined in GTR)\footnote{A manifold is a topological space. It is called an n-manifold if it is a Hausdorff-space, fulfils the axiom of countability, and is locally Euclidean.} takes the place of a universal in classical physics, while the instance of the magnitude, that which is represented by a measurement result (including the measurement error), is real. Thus, magnitudes in modern physics can be understood only via mathematics, which means: through the use of models. 

Magnitudes are in every case, in classical as in modern physics, associated with material entities. This means that magnitudes are either (i.) continuant magnitudes\footnote{A continuant is an entity that persists, endures, or continues to exist through time while maintaining its identity \cite{bfo:2020}.} (such as mass or density), which are \x{qualities} of material bodies; or they are (ii.) process magnitudes,\footnote{Processes are occurrents, in BFO terms, which means that they are entities that unfold through time, in other words entities that have temporal parts (parts along the temporal dimension).} such as velocity or force, where the processes measured involve material bodies as participants and interactions mediated by particles (as for example photons mediate the electromagnetic force).\footnote{Only in the case of gravitation are we unable to identify a material particle mediating the interaction (the graviton is a hypothetical particle).}

Each process magnitude at the instance level is what we call a process profile \cite{smith:2012c}.  This means that it is a chain of process characteristics, for example a chain of values of the expression $\ddot{x}\vert_{t=j}$, for the acceleration of a mass determined at different time points $j$.\footnote{An example is an ECG monitor used on a patient in an ICU. The electric currents of the cardiac conduction system visualized as lines on the monitor are such process profiles.} 

In physics, magnitudes are also called `physical quantities'. Here, however, we distinguish for clarity's sake between `magnitudes' on the one hand, and `quantities' on the other. We define a magnitude as an entity which can be quantified using measurements. Magnitudes are then of two sorts: (1) \textit{qualities} of continuants (for example \x{mass}, \x{length}), or \textit{characteristics} of processes (for example \x{velocity}, \x{acceleration}).\footnote{A quality, in BFO, is defined as `a specifically dependent continuant that, in contrast for example to dispositions, does not require any further process in order to be realized' \cite{bfo:2020}. Only continuants may have qualities, in BFO, and the term `characteristic' is used as the counterpart of 'quality' for occurrents.}
The magnitude is something repeatable that we find in reality in many particulars -- in classical physics, again, it is what was traditionally called a universal.\footnote{Magnitudes are assigned in each case to some particular. We can also say that a magnitude is a dimension of a phase space, namely of the phase space which models the behaviour of the underlying particular in the modelled system.}

A physical \x{quantity}, in contrast, is the count of how many of this or that particular magnitude are found in a specific case of measurement. It is that about which we speak when we collect measurement data and express it in terms of, for example, numbers (quantities) of meters or joules.  
The term `quantity' thereby captures also how much of or how many of a particular measurement magnitude we have in a system we observe or on which we perform measurements. In physics, the quantity is expressed using a mathematically defined magnitude that can be referenced using a variable in an equation. 

Physicists \textit{model} the real magnitudes encountered in nature by means of mathematical structures which are models of natural processes. 
To see what this means, consider an equation such as Newton's second law:

\begin{equation} \label{eq:n2}
F =  m \ddot{x} = \frac{md^2x}{dt^2}
\end{equation}

We have terms on the left, the middle and the right hand side of this double equation. A \x{mathematical term} is a meaningful (syntactically and semantically valid) combination of numbers, variables and symbols  that is used to designate structural entities in mathematics. An analogy in natural language is a noun or noun phrase in a sentence. There can be no false terms, but there are invalid terms, such as `$\frac{1}{0}$' or `$1\cdot$'. 

The referring expressions of the equation, `$F$', `$m$', and `$\ddot{x}$', play a dual role. \x{First,} they may refer to real magnitudes, in this case to the force, mass and acceleration, of some specific system element (a particle or body with a mass), thus describing an aspect of the reality of this element. When measurements of a system described by equation (\ref{eq:n2}) are made and the results plugged into the variables
of the equation, a law of nature is applied to particular measurements and the calculation prescribed by the equation is performed to yield a calculation result, for example in the form of a vector or -- in the above case -- a scalar with a unit of measurement.
The set of permissible values depends on the context in which the variable is used. For example in the term $\frac{1}{x},~ x \in \mathbb{R}\setminus{0}.$
 
\x{Secondly,} each variable of equation (\ref{eq:n2}) also represents a certain non-instantiable, abstract mathematical entity on the side of the human-created model itself. This second nature of the expression becomes evident when physicists manipulate equations using mathematical operations without using the equation with measurement results to calculate a value. The equation itself, and the referring expressions within it, are mind-dependent structures that are used by physicists to refer to measurable features of reality that exist independently of the human mind and of any measurement process.

\subsubsection{Quantification via constants}\label{const}

Constants, for example the gravitational constant, are magnitudes with a quantity that is fixed in a certain context.
Universal constants are thought to have fixed quantities in the entire universe. 

An example of a universal constant is the reduced Planck constant, measured in joule seconds, which is a continuant quality constant which captures the relationship between the energy of a photon and its frequency, $\hslash \approx 6.582 \times 10^{-16}eV\cdot s$. An example of a  universal process characteristic constant is the speed of light $c = 299792458 \frac{m}{s}$.

\subsubsection{Quantities and units of measurement} \label{uom}

Magnitude quantities -- the results of acts of measurement -- are expressed by means of a count together with a measurement unit. Occasionally we have pure counts (the mole\footnote{While the kilogram and metre are examples of measurement units used to quantify \textit{continuous quantities}, the mole is a measurement unit that is used for counting \textit{discrete quantities}. The mole is in fact not a unit of measurement, but rather a dimensionless counting measure -- a standardization of a count of particles -- introduced into the SI system for reasons of counting convenience. It appears in expressions such as `2 mol' or `2.00175 mol', where the latter is still, appearances notwithstanding, a discrete count (natural number)  $\in \mathbb{N}$, because $2.00175~\mathrm{mol} = 2.00175 \times 6.02214076 \times 10^{23}$ particles.}). 
A measurement unit is a fiat entity, which is a sub-entity of the corresponding quantity intuitively resulting from carving out equal divisions along a scale. It is itself a kind of quality or characteristic.  Measurement units are used in quantifying instances of the corresponding magnitudes \cite{johansson:2011}. 

The act of measuring consists in identifying the quantity of a physical magnitude (in the simplest cases by answering the question: \textit{How many}) of a given quality such as length or mass with either its discrete number of instances using natural numbers ($\mathbb{N}$) or using rational numbers ($\mathbb{Q}$) to approximate real numbers $\mathbb{R}$.

Leaving the mole aside, measurements resulting from use of measurement units are expressions which consist of two parts referring respectively to: 

\begin{enumerate}
\item[(i)] a rational number approximating a real number, and 
\item[(ii)] the unit of measurement itself, 
\end{enumerate}

\noindent joined together in expressions such as `$4.449$ kg'. 

That rational numbers (symbol: $\mathbb{Q}$) are used as approximations reflects practical constraints, including limits on precision of our measurement instruments, on the size of a display or on the storage capacity of a computer.\footnote{This is the reason why there are the floating point data types, which are formulaic representations of real numbers.} 
Each number, wherever it appears in a model in physics, whether $\mathbb{N}$, $\mathbb{Q}$, $\mathbb{R}$ or any other type of number (including complex numbers in $\mathbb{C}$), is a mathematical entity.  This is so whether the number in question is the result of a counting or measurement process or is inferred from an equation.\footnote{Many numbers in $\mathbb{N}$, $\mathbb{Q}$, $\mathbb{R}$ have real-world counterparts (for example the number of coins in my pocket is 3, the length of the hypotenuse drawing of a right-angled triangle with two sides of length $\approx 1$ is $\approx \sqrt{2} \in \mathbb{R}$). This is however not the case for numbers in $\mathbb{C}$.} 

\subsection{Models} \label{models}

A model in physics is a mind-dependent representation of an aspect of reality using abstract symbols that is created to describe, explain, or predict the aspect of reality in question. Abstract symbols are used primarily in the context of equations, but they may be accompanied by textual descriptions which may in turn be complemented by graphical, often geometrical, representations. 

The subject of the model is either a system, a system element, or an interaction. 
Examples are: for system elements: an electron; for systems: the solar system, or the two-element hydrogen atom system consisting of a proton and an electron; and for interactions: gravity, in the case of the solar system; or the electromagnetic force through which, in the case of the hydrogen atom, proton and electron interact.
When two systems interact, they are modelled as system elements on a more coarse-grained level. This is how the granularity levels of systems treated by physics range from elementary particles to galaxies.
The equations in the model represent the model's subjects by means of mathematical entities. The equations describe the relationships between these entities using mathematical relations. They thereby model physical reality in an idealised form. Therefore, models are not exact representations of associated system entities in reality. They merely approximate such entities. Just as real-world shapes are approximated by mathematical shapes, so real-world processes are approximated by the models of physics. 

In classical physics, we have systems in reality which are modelled using universals, magnitudes, which have individual instances but are represented in the models as mathematical entities.
\x{In modern physics, in contrast, we have system entities, magnitudes and models which contain no universals, but only mathematical entities} (see section \ref{ontorel}).
We now turn to the mathematics upper ontology needed to describe the formal models in physics.

\section{A mathematics upper ontology} \label{matho}

\subsection{The nature of mathematical entities}

Mathematical entities are ideal, which means that they are not part of the causal world of time and change. They are intrinsically intelligible entities (which means that propositions about them can be known \x{a priori}). Further, they are mind-dependent; but at the same time they are independent of our sensory experience of the world outside our mind \cite[pp.~69f.]{reinach:1989}. 

We saw that they are not universals in anything like the Aristotelian sense in which we use this term, and they are also not instances, and nor do they have instances, unless, as we shall see in section \ref{epm}, instances are assigned. In physics, chemistry and also in computer science and other branches of applied mathematics, mathematical entities are used to model reality; but this does not mean that there is something in reality to which these models directly correspond. 

The domain of mathematical entities ranges from very simple examples such as numbers and basic shapes, which we appeal to when performing acts of counting or of describing real entities such as tabletops or pieces of string, to highly abstract entities such as Hilbert spaces or Lebesgue integrals, which can be understood only with the aid of equations.

\subsection{Mathematics upper ontology}

There are three types of mathematical entities: monads, constructors and structures. Monads are atomic primitives from which structures are made through the application of constructors.\footnote{Our ontology of mathematics differs from mathematical structuralism, for example as described by \textcite{reck:2020}, because we postulate, like Frege, that there are primitives (monads) which are not merely determined by the context of other mathematical entities, but have an existence in their own right.} Examples of monads are zero, one, or the point.

Constructors are used to obtain structures from monads and structures. For example, the number $2$ is constructed with the equality constructor and the addition operator (a constructor) like this: $1+1=2$.\footnote{Note that the monad $1$ is here used twice.} 
Simple operators\footnote{An example of an operator in the sense of functional analysis is discussed in section \ref{proj}.} are constructors used to perform basic mathematical operations such as addition and multiplication. Their relata are determined by the axioms of mathematics (for example, the Peano axioms) chosen.
Structures in mathematics are essential structures, which means that they are entities whose component parts stand in necessary relations to each other. We are influenced here by the German philosopher Adolf Reinach, who spoke of `essential connections' \textit{(Wesenszusammenh\"ange)} thereby defending the idea that there is a wide class of material necessities which can be known a priori in domains such as color and shape, rational psychology, and above all social ontology. Reinach's uniqueness consists in the fact that he was one of the first to recognize essential structures not only in the timeless realm of mathematics \textit{\`a la} Plato, but also in the historically changing realm of debt and ownership.\footnote{There is an essential connection, for example, between a promise, a claim and an obligation. The latter exist as nodes in a system of necessary connections alongside other nodes such as intention (to act as promised), realization (of the promise), waiving (of the claim) and so forth \cite{smith:1992}.}
A helpful example of an essential structure in the Reinachian sense, slightly more involved than $1+1=2$ is the constant structure $\pi$.\footnote{\label{pi}$\pi$ is a structure that is obtained from the monad `$1$' by first defining the structure \x{plane} ($E=\{x,y,z \in \mathbb{R}|ax+by+cy=d\}$) using the constructors $=, \in$, monadic variables and the set constructor as well as the addition and multiplication operators and distance (a structure defined as the segment of the line $\alpha x + \beta y = \gamma$) and then by constructing the circle as the set $\{X \in E| \overline{MX} =r \}$ in the plane $E$ with central point $M \in E$ and a distance $r\subset E$ called radius with circumference $C$. $\pi$ is then defined as $\pi=\sfrac{C}{2r}.$}

In mathematics, all of these entities are intrinsically intelligible (= \x{a priori}) entities which have no instances in reality.\footnote{Other Reinachian \x{a priori} structures, such as the promise, have instances.}  Using constructors and monads, all mathematical entities can be obtained.

\begin{figure}[htb]
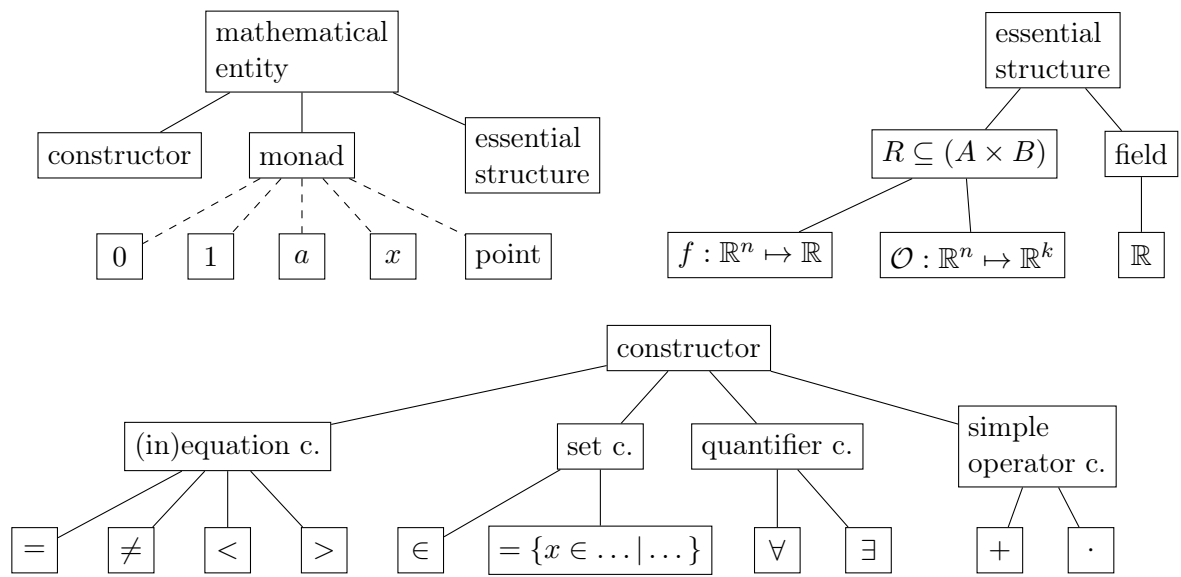

\begin{center}

\begin{tabular}{cc}

\begin{tabular}{c}
\begin{minipage}[t]{.45\linewidth}
\begin{tikzpicture}
  \tikzstyle{e}  = [>=latex]
  \tikzstyle{ne} = [e,dashed]
  \tikzstyle{t}  = [draw,rectangle,minimum height=6mm,minimum width=6mm]

\input{ontoPhysics-dot2tex-fig1}
\end{tikzpicture}
\end{minipage}
\end{tabular}
\hspace{20pt}
& 
\begin{tabular}{c}
\begin{minipage}[t]{.45\linewidth}
\begin{tikzpicture}
  \tikzstyle{e}  = [>=latex]
  \tikzstyle{ne} = [e,dashed]
  \tikzstyle{t}  = [draw,rectangle,minimum height=6mm,minimum width=6mm]
\input{ontoPhysics-dot2tex-fig2}
\end{tikzpicture}
\end{minipage}
\end{tabular}
\\
\end{tabular}

\bigskip

\begin{tikzpicture}
  \tikzstyle{e}  = [>=latex]
  \tikzstyle{ne} = [e,dashed]
  \tikzstyle{t}  = [draw,rectangle,minimum height=6mm,minimum width=6mm]

\input{ontoPhysics-dot2tex-fig3}
\end{tikzpicture}

\caption{Top-level entities of mathematics with examples of entities at lower levels. $A \times B = \{(a,b) \mid a \in A, b \in B\} $, $R$: relation with relata (domain and range), $\times$: cross product, $\mathbb{R}$: real numbers. $\forall$ and $\exists$ are fully expressed as constuctors like this: $ \exists x_1,  \dots , x_n | \dots ,$ where the second `$\dots$' indicates a syntactically correct term of predicate logic obtained using zero or more of the connectives $\lnot, \lor, \land$ or $\to$ and using the variables $x_1, \dots , x_n, n \geq 1,$ possibly with n-ary predicates $R_1, \dots , R_m$ as well as constants $a_1, \dots , a_\ell$. $f$ -- functional, $\mathcal{O}$ -- operator (functional analysis). Note that the solid edges of the graph mean $\subseteq$, whereas the dotted edges mean $\in$, so that the monads, which are primitive elements not regarded as sets here, are shown as elements of the set of monads (and not as subsets). In the constructor taxonomy at the bottom `c.' means `constructor'. \label{f:math}}
\end{center}
\end{figure}

Figure \ref{f:math} shows some important constructors with their associated monads and essential structures. The latter are always sets in the mathematical sense.\footnote{Or in some cases they are classes in the Neumann-Bernays-G\"odel (NBG) axiomatization of set theory \cite[ch.~4]{mendelson:1997}.}
Therefore, unlike the ontology of physics, in which the edges represent the `is a' relation, in the mathematics ontology, it signifies `subset of' ($\subseteq$): The entire realm of mathematical entities is built up from monads (0 and 1, the geometrical point, etc.) using set theory, and all the relations in the ontology are set-theoretical (using the constructors). As in mainstream mathematics, we assume a set theory (in almost all cases using Zermelo-Fraenkel axiomatization with the axiom of choice (ZFC) will suffice for our purposes \cite{ebbinghaus:1992}). For example, the type $R \subseteq A \times B$ ranging over the sets $A$ and $B$ as its relata is a superset of one of its subset types, the set of functionals $f$ (functional is the designation of function in functional analysis). The type $R$ is also the superset of the type operator (in the sense of functional analysis), an example of which is $\nabla$, the gradient operator, or its special case, the univariate differential $\frac{d}{dx}$.

Note that we do not discuss the \x{expressions} we use to formulate mathematical models. We are interested in the models themselves. When we use expressions such as `equation', we are referring not to a string of symbols but rather to mathematical entities which the symbols represent. Using examples in the next section, we will see how mathematical entities can be classified using the upper ontology shown in Figure \ref{f:math}.

\section{Examples from classical and quantum physics}
We proceed with an ontological analysis of one case from classical physics, the harmonic oscillator, and another from quantum physics, the entangled photons.

\subsection{The harmonic oscillator} \label{ho}

The simple harmonic oscillator is a purely idealized model which is used as a template for further models which successively approximate real oscillators more closely, for example the harmonic osciallator with dampening used to model electric circuits. In the simple harmonic oscillator, a mass that is displaced from its equilibrium position undergoes a restoring force  proportional to the displacement and undergoes changes of distance along only one dimension.

An example for such an idealized, imaginary system -- imaginary not least in that friction is entirely ignored -- is a weight suspended from a spring that is subjected to displacement by means of an imagined force. Here the spring provides a restoring force that is proportional to the displacement  of the weight, as defined in the differential equation 

\begin{equation} \label{eq:osc}
 \frac{md^2x}{dt^2} =  - kx, 
\end{equation}

Here $m$ is the mass of the oscillating weight, $t$ is the time interval defining the acceleration of the mass which constitutes the exerted force, $x$ is the displacement distance defined as a function of $t$ (so `$x$' would be written out in full as `$x(t)$'), and $k$ is the constant of the retraction force, also called `stiffness'.

The equation is part of a model that represents an imaginary harmonic oscillator, whose movement is assumed to be without friction. This means that its total energy is constant, and when averaged over time (over one period, $2\pi$) the kinetic energy equals the potential energy. 


On the left-hand side of the equation, we have a representation of the force $F$ as defined by Newton's second law (see equation (\ref{eq:n2}) above). 
On the right-hand side, we have a representation of this same force exerted by the spring.\footnote{We note in passing that the minus sign on the right-hand side denotes the opposing directions of the two forces this equation describes. We note in passing, too, that we are viewing the ideal harmonic oscillator as an inertial frame of reference, that is, as a system imagined to be moving at a constant velocity. We make this assumption in order to avoid the need to enter into any relativistic discussion of spacetime.}


The solution to equation (\ref{eq:osc}) is 

\begin{equation}\label{eq:osol}
x = a \cos(\omega t + \Delta),
\end{equation}

with constant amplitude $a$ (the maximal extent of change over a period), angular frequency $\omega = \sqrt{\frac{k}{m}}$, and constant phase $\Delta$, the starting point of the periodic function.

The values of $a$ and $\Delta$ depend on with how much force the motion\footnote{Recall that we are here dealing with idealized magnitudes and system entities only.} is initiated.\footnote{Once initiated it remains constant -- recall that this is an idealized model.} The value of $\omega$ depends on the properties of the idealised oscillator itself (namely on its mass and on the retraction force).

With these variables and this solution, we can \x{calculate} the idealised position $x$ of the oscillating mass at any time $t$ in an imaginary phase space. A phase space is the algebraic field which is used by the model of the system to obtain the required model entities which are elements of or defined over the field.

\subsubsection{Mathematical ontology of the oscillator model}

Like all models (in the sense of this term that concerns us here), the oscillator model consists of equations, text and (optionally) figures. This model is built to represent a physical system, albeit one that is imaginary. But it can also be viewed as representing a mathematical structure. 
We discuss ontologically only the linear differential equation (\ref{eq:osc}) here.
As we saw, it has the form $\frac{md^2x}{dt^2} =  - kx$.
This equation asserts the equality (which is created using the constructor $=$, see figure \ref{f:math}) 
of two essential structures: (1) the product of the mass variable $m$ with the double derivative relative to the time variable $t$ of the distance function $x(t)$, and (2) the negative of the product of a constant $k$ and the distance function $x$. The negation and product are operators.
 $x(t)$ is a function $f:\mathbb{R}\mapsto \mathbb{R}$ (see Figure \ref{f:math}). Its derivative is an essential structure, a binary relation from the tangent space\footnote{The tangent space $T_{x}M$ of a manifold $M$ is the set of all tangential vectors $v={\frac{d\gamma }{dt}}(0)\in T_{x}M$  with the differentiable curve $\gamma(0)=x$ and curve parameter $t$ for all points $x \in M$. This, too, is a structure in MathO, but we do not analyse it further here.} of a functional to $\mathbb{R}$, and this is also true for the double derivative $\frac{d^2}{dt^2}$. Both yield a scalar.

Because the \x{idealized} harmonic oscillator only moves up and down in one direction, vectors are not needed to model its motion.
The phase space which is required for the model is just the simple space of real numbers $\mathbb{R}$, an algebraic field.

\begin{figure}[htb]
\begin{center}
\begin{tikzpicture}
  \tikzstyle{e}  = [>=latex]
  \tikzstyle{ne} = [e,dashed]
  \tikzstyle{t}  = [draw,rectangle,minimum height=6mm,minimum width=6mm]
\input{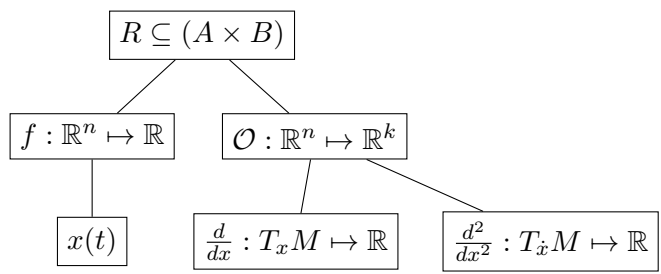}
\end{tikzpicture}
\caption{Mathematical relations: Functionals and operators. \label{mr}}
\end{center}
\end{figure}

Because of the set-theoretic relationships which define the taxonomy of mathematical entities, the ontological structure is shown in Fig. \ref{mr} (all edges mean $\subseteq$).

We will see in the quantum physics example below a radically more complicated and also more abstract (MathO) modelling space.

\subsubsection{Magnitudes in the harmonic oscillator model} \label{m:phot}

The harmonic oscillator model (\ref{eq:osc}) contains four MathO entities, which in turn represent four PhysO magnitudes. Note that because all mathematical entities other than monads are sets, magnitudes mapped to mathematical entities in physics equations are sets.

The magnitudes are part of the system that is represented by the physics and mathematics entities: mass ($m$), distance ($x$), time ($t$), and acceleration ($\ddot{x}$) and the spring constant (stiffness) $k$. 
Mass, distance and the spring constant are continuant magnitudes. \x{Mass} is a quality of a body that `does not require any further process in order to be realized' \cite{bfo:2020}.  The \x{distance} $x$ is a length, a primitive quality in our common sense understanding of the world that is also a mathematical entity, see the definition of $\pi$ in footnote \ref{pi}. The \x{retraction} (spring) constant $k$ is the quality of a spring which accounts for its retraction force.

\begin{figure}[htb]
\begin{center}
\begin{tikzpicture}
  \tikzstyle{e}  = [>=latex]
  \tikzstyle{ne} = [e,dashed]
  \tikzstyle{t}  = [draw,rectangle,minimum height=6mm,minimum width=6mm]
\input{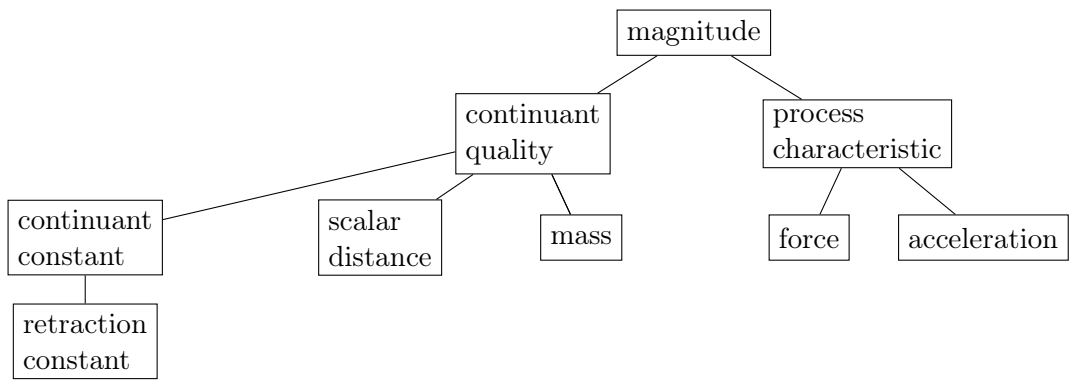}
\end{tikzpicture}
\caption{Harmonic oscillator magnitudes. \label{harmo}}
\end{center}
\end{figure}

These magnitudes have the ontological relationships shown in Fig. \ref{harmo}.

The acceleration referred to on the left hand side of equation (\ref{eq:osc}), $\ddot{x}=\frac{d^2x}{dt^2},$ is a measurable process profile\footnote{For an  account of the term `process profile' see BFO 2.0 \cite{smith:2012c}. To say that velocity and acceleration are process profiles is to say that they are proper parts of some (intuitively) larger process of motion. They are that part upon which our focus is directed when we carry out the corresponding measurement.} of a process of motion. It  depends on the motion and can exist only if the motion exists. Without motion, there is no acceleration.  Acceleration is a process magnitude.  

The force $F$ shown in equation (\ref{eq:n2}) is also a process magnitude. It is the effect on a body (mass) which it accelerates, which means that it changes the amount or direction of velocity of the moving mass or deforms its body. `Process' is defined in such a way that all processes have temporal parts. This applies to acceleration also, which is a derivative of velocity with respect to time,\footnote{Or spacetime, in the language of the general theory of relativity.} and so has temporal parts also from its mathematical definition. When acceleration is measured at an instant, then a measurement result, a continuant magnitude, is obtained. All such results are merely approximations to their real-world counterparts.

What, now, does an ontological analysis of the equation for the harmonic oscillator (\ref{eq:osol}) and its solution 
yield? From the perspective of physics, the model describes the relationships between physical magnitudes which are used to model 
natural systems.
When experiments are performed, these magnitudes are measured, and units of measure have been invented for this purpose. 
But the way the magnitudes relate to each other is described using mathematical equations. 

From the perspective of mathematics, the magnitudes used in the models are mathematical structures. From the perspective of physics, they describe properties of system elements and their interactions.\footnote{An overview is given in figure \ref{f:venn}.}

\subsection{Entangled photons model} \label{epm}
We now consider a very simple, highly artificial (but experimentally realised) quantum system consisting of two particles (entangled low-energy photons). We use this system as our example, because it enables us to discuss many important ontological questions relating to the boundary between classical and modern physics. We believe that an ontology that can cope with this system can be used for any system in the quantum domain. It can be used also for the general theory of relativity, because there, too, the main problem is an adequate representation -- and provides an understanding -- of its mathematical components.

A photon is simply a system element in the physics ontology, a particle.  The existence of an entangled system is commonly seen as proving that quantum physics leads to the metaphysical conclusion that there are non-local dynamic effects in nature, or in other words that there is ``action-at-a-distance.'' \cite[p.~486]{maudlin:2003} To see whether this conclusion follows, we now analyse this type of system ontologically.

Using a technique called `spontaneous parametric down-conversion' one single high energy photon of \x{spin one} can be converted into a pair of entangled low-energy photons $a$ and $b$ each of \x{spin half}, as outlined in \cite{boyd:2020}.
The photons obtained in such experiments are highly artificial, in the sense that a process of the given sort can be realized only in very special artificially contrived circumstances. 

Parametric\footnote{The process is called parametric because its underlying quantum effect can be modelled using a parametric (exponential family derived) non-linearity.} down-conversion ``is a nonlinear instant optical process that converts one photon of higher energy (namely, a pump photon), into a pair of photons (namely, a signal photon, and an idler photon) of lower energy, in accordance with the law of conservation of energy and the law of conservation of momentum.''\footnote{Source: https://en.wikipedia.org/wiki/Spontaneous\_parametric\_down-conversion.} 

The spin \label{spin} is an immutable inner quantum property of a particle that has no universal. Rather, in place of the universal, we have a mathematical entity (see page \pageref{declared}). Nevertheless, we can perform a measurement of spin experimentally, for example by exploiting the real magnetic moment caused by it, an occurent magnitude. When we say that we measure spin, we declare that the measurement is related to certain mathematical entities we have decided to use in building up quantum mechanics. Spin has the characteristics of the classic angular momentum, namely it satisfies a conservation law and it can undergo geometric transformation. Yet, it cannot be explained as the rotation of a mass. That is just a pretty picture. 

When we say that the two particles are entangled, what we mean is that the spin of the particles is complementary, in the sense that if $a$ is spin up in $z$-direction, then $b$ is spin down in that direction and vice versa. Because the particles are in a state of superposition, which one is spin up and which one down is not known before a measurement on one of the respective particles is performed. This state can be modelled using the singlet\footnote{Singlet because the entangled photons resulted from one single photon and conserve its energy although they are separated in space.} wave function:

\begin{equation}\label{e:sup}
 \ket{\phi_0} = \frac{1}{\sqrt{2}} \left(\ket{z_a^+ z_b^-} - \ket{z_a^- z_b^+}\right),
\end{equation}

where $\ket{z_a^+ z_b^-}$ is a quantum state in which particle $a$ has the physical property $S_{az}=\sfrac{+1}{2}$, particle $b$ has the physical property $S_{bz}=\sfrac{-1}{2}$, and $\ket{z_a^- z_b^+}$ has the analogous meaning with the opposite sign. The qualities $S_{az}$ and $S_{bz}$ are the spin quantities of the particles in the \x{z}-direction. In MathO, they are classified as projectors (see \ref{proj}). The equation says that these properties are superposed in opposing particle states (the particle state at a point in time, i.e. the set of measurable values of the particle's non-invariant properties at a point in time).
To understand what this means, we need to analyse the ontologies of mathematics (MathO) and of magnitudes (PhysO) as they apply in this entangled particle model, to which we now turn. 

\subsubsection{Mathematical ontology of the entangled photon model}

In the entangled photon model formed by equation (\ref{e:sup}), the following entities appear which are not present in the harmonic oscillator model (eqn. (\ref{eq:osc})):

\begin{enumerate}
\item[(1)] fraction and square root. These are operators like multiplication or addition.
   \item[(2)] parentheses which are paired precedence operators used to change or highlight precedence in equations. Here they allow the distributive usage of the factor $\sfrac{1}{\sqrt{2}}$ to simplify the appearance of the expression. 
\end{enumerate}
   
\begin{figure}[htb]
\begin{center}
\begin{tikzpicture}
  \tikzstyle{e}  = [>=latex]
  \tikzstyle{ne} = [e,dashed]
  \tikzstyle{t}  = [draw,rectangle,minimum height=6mm,minimum width=6mm]
\input{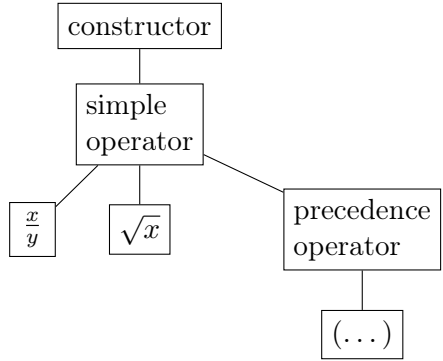}
\end{tikzpicture}
\caption{Simple operators \label{siop}}

\end{center}
\end{figure}

   
Given the set-theoretical nature of the mathematics ontology, we have the ontological structure shown in Fig. \ref{siop}.

But there is also the Dirac notation $\ket{\dots}$, which is used to denote a state (vector) in a finite Hilbert space $\mathcal{H}$ as follows:\footnote{Chapters 1 to 6 of \cite{griffiths:2002} give an excellent introduction to the mathematics of quantum mechanics on which our account here is based.}

\begin{equation} \label{e:base}
\ket{\psi} = \sum_j \ket{j}\bra{j}\ket{\psi},
\end{equation}

so that the state vector $\ket{\psi}$ is the sum of all the amplitudes (a quantum mechanics specific pre-probability defined in next paragraph (eqn. \ref{e:ip})) of state $\ket{\psi}$ to be in each of the base states $\ket{j}$ of $\mathcal{H}$ multiplied by $\ket{j}$.\footnote{For an introduction to quantum mechanics, see volume 3 of \cite{feynman:2010}; Dirac notation is explained in detail in \cite{griffiths:2002}, chapters 3-4, 6-7. } 

$\mathcal{H}$ is equipped with an inner product \cite[ch.~3-4]{griffiths:2002} between two wave function vectors $\ket{\phi}$ and $\ket{\psi}$ given by

\begin{equation}\label{e:ip}
 \mathcal{I}(\ket{\phi}, \ket{\psi})= \bra{\phi}\ket{\psi} = \sum_m \phi^*(m)\psi(m),
\end{equation}

where the $m$ are the base vectors of a finite Hilbert space (see paragraph \ref{hilbert} and \cite[ch.~3]{griffiths:2002}). This inner product is used in physics to express the probability \x{amplitude} of state $\ket{\psi}$ to move to state $\ket{\phi}$. It expresses a relationship between the two states as a probabilistic measure in the form of a complex number.\footnote{The probabilistic view of quantum mechanics has been formalised thoroughly, for example in \textcite[ch.~5]{griffiths:2002}. We do not show the ontological representation of this formalism here, but it can be obtained in a straight-forward manner.}
The square of its absolute $|\bra{\phi}\ket{\psi}|^2$ is the corresponding quantum probability (Born rule).
For example, the inner product $\mathcal{I}(\ket{\phi}, \ket{\psi})$ (eqn. \ref{e:ip}) is used to express the amplitude of a particle in state $\ket{\psi}$ with a momentum $p$ to be found at a position $x$ as 

\begin{equation}
 \bra{x}\ket{\psi} = \psi(x) \propto e^{\sfrac{+ipx}{\hslash}},
\end{equation}

which is a complex number. 

Here we have the constructor $\propto$ which expresses proportionality, the dependence of one magnitude (output) upon one or more others (input)\footnote{Proportionality is ubiquitous in physics, for example the harmonic oscillator shown in eqn. (\ref{eq:osc}) has the proportionality relation $F \propto -x$.}, the exponential function, a relation, and the imaginary number $i = \sqrt{-1}$.

The denominator of the exponential term is the reduced Planck constant $\hslash$ which we encountered in section \ref{const}. From the perspective of mathematics, the Planck constant is just a scalar.

How are these entities treated ontologically?

\paragraph{Hilbert space} \label{hilbert}

A Hilbert space $\mathcal{H}$ used in quantum physics is a vector space over the field of complex numbers $\mathbb{C}$ endowed with an inner product (shown in eqn. (\ref{e:ip}) above, and in \ref{inner}) that turns it into a metric space. As a metric space, it is complete with respect to the norm induced by the inner product, 
i.e. every Cauchy sequence\footnote{A sequence in which the distance of its elements 
shrinks arbitrarily as the sequence progresses.} of points in $\mathcal{H}$ has a limit that is also in $\mathcal{H}$. In many highly idealised models of physics, it is conceived as finite, but there is an infinite Hilbert space for quantum mechanics as well, which was conceived by Dirac. This conception is problematic, however, because to define it one must use the Dirac function, whose value is zero everywhere except at zero, yet whose integral over the entire real line is equal to 1. 

The problem is that a function cannot have an integral of 1 on an infinitely small domain interval.\footnote{There is a derivation of the Dirac function as a linear form acting on functions in the theory of distributions (L. Schwartz).} So this function is mathematically irregular (it cannot be represented as a locally integrable function $\delta(f) = \int_\Omega \delta(x) f(x) dx = f(0), \Omega \subset \mathbb{R}^n$). But it is needed in order to maintain the linear independence of the states in an infinite-dimensional Hilbert space. 
We have here, then, an important example of the limits of mathematical modeling of nature.

For a finite Hilbert space $\mathcal{H}$, an orthonormal state basis of non-zero kets $\{ \ket{\phi_1}, \ket{\phi_2}, \dots \}$ can be defined so that $\forall j,k~\mathsf{with}~j \neq k | \bra{\phi_j}\ket{\phi_k}=0$ (all kets are orthogonal) so that

$$
\bra{\phi_j}\ket{\phi_k} = \delta_{jk}.
$$

\begin{figure}[htb]
\begin{center}
\begin{tikzpicture}
  \tikzstyle{e}  = [>=latex]
  \tikzstyle{ne} = [e,dashed]
  \tikzstyle{t}  = [draw,rectangle,minimum height=6mm,minimum width=6mm]
\input{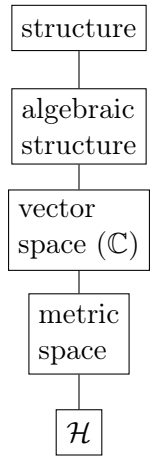}
\end{tikzpicture}

\caption{Hilbert space in ontological context. \label{ha}}
\end{center}
\end{figure}

Given that the edges in the mathematics ontology are subset-relations, $\mathcal{H}$ can be represented as shown in Fig. \ref{ha}.

Note that in quantum physics the Hilbert space is the phase space of the modelled system.

\begin{figure}[htb]
\begin{center}
\begin{tikzpicture}
  \tikzstyle{e}  = [>=latex]
  \tikzstyle{ne} = [e,dashed]
  \tikzstyle{t}  = [draw,rectangle,minimum height=6mm,minimum width=6mm]
\input{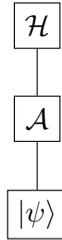}
\end{tikzpicture}
\caption{Quantum state vector in ontological context.\label{ket}}
\end{center}
\end{figure}

\paragraph{State vector}
An element of a finite $\mathcal{H}$ is a quantum state vector $\ket{\psi}$ which can be expressed as a linear combination of the basis vectors $j$ of $\mathcal{H}$ as indicated in equation (\ref{e:base}). One or more state vectors can form a subspace $\mathcal{A} \subseteq \mathcal{H}$.
Ontologically we have the structure shown in Fig \ref{ket}.

\paragraph{Inner product}\label{inner}
An inner product (the operator is written as $\cdot$) is a relation which assigns a scalar to two vectors, for example: $\mb{W}=\mb{F}\cdot\mb{s}=|\mb{F}||\mb{s}\|\cos \phi,$
where $\mb{W}$, $\mb{F}$ and $\mb{s}$ are the work, force and distance vectors, respectively, and $\phi$ is the angle between the force and the distance vector.

In a quantum Hilbert space, the inner product $\mathcal{I}(\ket{\phi}, \ket{\psi})$ defined in equation (\ref{e:ip}) yields a complex number indicating the amplitude of a particle to change from one state into another.

\begin{figure}[htb]
\begin{center}
\begin{tikzpicture}
  \tikzstyle{e}  = [>=latex]
  \tikzstyle{ne} = [e,dashed]
  \tikzstyle{t}  = [draw,rectangle,minimum height=6mm,minimum width=6mm]
\input{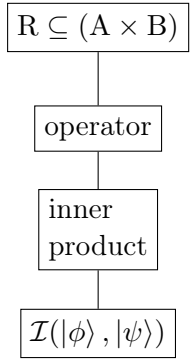}
\end{tikzpicture}
\caption{Inner product in ontological context. \label{inp}}

\end{center}
\end{figure}

Ontologically, we have the relationships shown in Fig. \ref{inp}.

\paragraph{Projector}\label{proj}

The spin quantities of the particles $a$ and $b$ in equation (\ref{e:sup}) measured in the \x{z}-direction are $S_{az}$ and $S_{bz}$. In the mathematics of quantum mechanics they are modelled as Hilbert projectors, which are operators with properties defined as follows.

We first define an operator $A$ in quantum mechanics. It is a binary relation which maps a ket (a vector subset of $\mathcal{H}$ used to model quantum mechanical states, see \cite[ch.~3]{griffiths:2002}) to another ket:

$$
A:\mathcal{H} \mapsto \mathcal{H}, A(\alpha\ket{\phi}+\beta\ket{\psi}) = \alpha A (\ket{\phi}) + \beta A \ket{\psi}
$$

with $\ket{\phi}, \ket{\psi} \in \mathcal{H}$ and $\alpha, \beta \in \mathbb{C}.$ Then a projector is an operator $P$ with the following additional properties:

$$
P^2 = P, ~~~~~~ P^\dagger = P,
$$

where $P^\dagger$ indicates the complex conjugate transpose\footnote{This is also known as the Hermitian transpose: For any $m \times n$ complex matrix $A$, this is an $n \times m$ matrix obtained by transposing $A$ and applying the complex conjugate on each entry.} of $P$.
The $\ket{j}\bra{j}$ referred to in equation (\ref{e:base}) is also called a dyadic projector. 

\begin{figure}[htb]
\begin{center}
\begin{tikzpicture}
  \tikzstyle{e}  = [>=latex]
  \tikzstyle{ne} = [e,dashed]
  \tikzstyle{t}  = [draw,rectangle,minimum height=6mm,minimum width=6mm]
\input{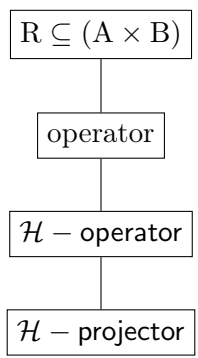}
\end{tikzpicture}

\caption{Quantum projectors in ontological context. \label{qpr}}
\end{center}
\end{figure}

Ontologically, we have the stucture shown in Fig. \ref{qpr}. 

\paragraph{Tensor product}\label{tens}

To understand the ontology of the entangled photon system we need to understand the tensor product $\otimes$. In the quantum mechanics formalism of Griffiths, a tensor product of two Hilbert spaces is used to describe two related systems which together form a new system, such as the two particles described in equation (\ref{e:sup}). This tensor product is defined as follows: Let $\ket{a_j}$ and $\ket{b_p}$ with $j=1 \dots m, p = 1 \dots n$ be orthonormal bases\footnote{An orthonormal basis is a set of linearly independent basis vectors of a vector space which are all normed to length one.} for the Hilbert spaces $\mathcal{A}^m \subset \mathcal{H}$ and $\mathcal{B}^n \subset \mathcal{H}$, respectively. Then the collection of $mn$ elements $\ket{a_j} \otimes \ket{b_p}$ forms an orthonormal basis of the tensor product $\mathcal{A}^m \otimes \mathcal{B}^n.$ This is the set of all linear combinations
 \begin{equation}
 \ket{\psi} = \sum_j \sum_p \gamma_{jp} (\ket{a_j} \otimes \ket{b_p}),
 \end{equation}
 where the $\gamma_{jp}$ are complex numbers \cite[sect.~6.2]{griffiths:2002}.

\subsubsection{Magnitudes in the entangled photons model}
The model defined by equation (\ref{e:sup}) contains only one magnitude, the spin, which can take the quantities $w^+ = \frac{\hslash}{2}$ and $w^-=-\frac{\hslash}{2}$, where $w$ is an arbitrary axis in $\mathbb{R}^3,$ usually $x, y$ or $z$. 

Here `$+$' and `$-$' represent spin qualitities referred to as `up' and `down', according to the way that they are measured.

The spin of a particle is the quantum mechanical equivalent of the angular momentum of classical physics, but with the restrictions made at the beginning of this section. 

But where the latter can be imagined using our common sense knowledge of the world, this is not so of the former, which can only be understood as a property that leads to certain indirect measurements results (see above page \pageref{spin}). 
We model this property mathematically. It is a vector magnitude because it has three spatial components $\mathbf{s} = (s_x, s_y, s_z)$. It is a vector modelling a conserved quality of the particle and is therefore a continuant quality like mass and not a process characteristic like acceleration.

\subsubsection{The meaning of the model}
From the ontological analysis of the entangled photon model, we can derive the following interpretation  of equation (\ref{e:sup}), which we repeat for convenience: $$\ket{\phi_0} = \frac{1}{\sqrt{2}} \left(\ket{z_a^+ z_b^-} - \ket{z_a^- z_b^+}\right).$$
In the notation of Griffiths $\ket{z_a^+ z_b^-}= \ket{z^+}_a \otimes \ket{z^-}_b$ is ``an eigenstate of both $S_{az}$ [z-direction spin] for the $a$ particle, eigenvalue $\sfrac{+1}{2}$, and of $S_{bz}$ for the $b$ particle, eigenvalue $\sfrac{-1}{2}$; the state $\ket{z_a^- z_b^+}$ has a similar interpretation with eigenvalues of the opposite sign.'' \cite[sect.~V~A]{griffiths:2011}.\footnote{In the linear algebra of the Hilbert space describing the quantum system under consideration here, an eigenstate and its eigenvalue are the results of a linear mapping of $\mathcal{H}$ on itself (an endomorphism).}
That the projectors $S_{az}$ and $S_{bz}$ have the tensor product $\ket{z^+}_a \otimes \ket{z^-}_b$ as eigenstate means that they have a physical interpretation in the model of the spin states of the photons with phase space $\mathcal{A}^m \otimes \mathcal{B}^n$.
One can also show how to add the experimental process of the measurement of the spin
to the phase space of the model \cite{griffiths:2011}.

But the superposition expressed as $\ket{\phi_0}$ in equation (\ref{e:sup}) is not an eigenstate of $S_{az}$ or $S_{bz}$, and therefore the state cannot be interpreted as a property of the system because the measurement operators cannot be related to it. 

Therefore equation (\ref{e:sup}) and the state it describes have no physical interpretation, and it makes no sense to say that a quantum system with property $\phi_0$ has any non-trivial property corresponding to a subspace of the Hilbert spaces $\mathcal{A}$ or $\mathcal{B}$  or their tensor product (see \ref{tens}). From this perspective, the entanglement \x{per se} does not mean that there are non-local effects. The experiments performed on entangled systems do, certainly, seem to imply this.\footnote{The consistent histories school \cite{griffiths:2002,omnes:1994} has argued that there is no such implication, but at the price of losing world unity, i.e. the view that we can understand the world as a coherent whole.
}
But we think that the ontological character of entanglement is in this sense void -- that there is nothing that we can learn from this entanglement model about the mode of existence of the world. We can, however, describe the ontological significance of various parts of the entanglement model. 

Each of the ket-terms on the right side of equation (\ref{e:sup}) has an ontological meaning in physics, namely opposite spin directions of the two particles. But $\ket{\phi_0}$ is ontologically void, we cannot imagine or understand it.

The equation describes a state superposition that is generated using a machine and that can be dissolved using a second (measurement) machine.
When the particles are prepared by the first machine, which creates the spontaneous parametric down-conversion of the high energy photon, we actively prepare a state which has no ontological significance in particle physics, but only the state of being engineered (a technical entity -- \textgreek{teqn'h| >'wn}, Aristotle, \x{Physics} $\Gamma$).
The entangled particles form a pattern (in information theory: contain information) that we can model mathematically using the singlet model of equation (\ref{e:sup}); but we cannot give it any ontological significance other than that these are artificial particles that show a characteristic pattern. 
  
When we measure the particles at different points in space, we do not get any non-local effects, but we just use another machine to recover the information we  had earlier introduced. While doing this, we learn about what we can achieve by applying sophisticated machines to natural particles. We therefore learn something that \x{is} ontologically valid, namely about our machines.

\section{The ontological relation of common sense, physics and mathematics} \label{ontorel}

\begin{figure}[htb]
\includegraphics[width=\textwidth]{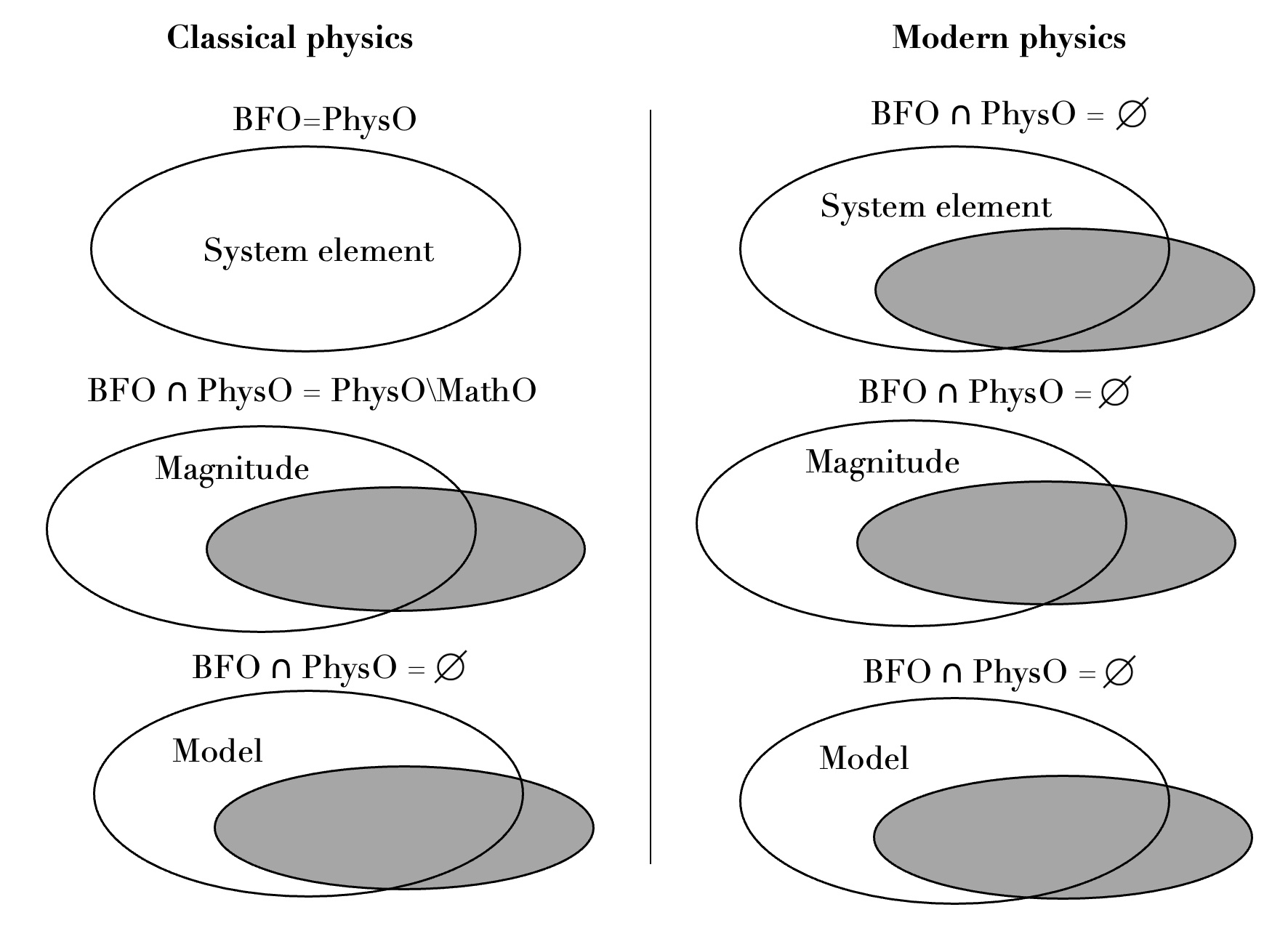}
\caption{Ontological relationships between common-sense ontology (BFO) and the physics and mathematics ontologies (PhysO and MathO) in classical and modern physics. PhysO: open set, MathO: filled set (grey). Relations between system element, magnitude and model not shown. \label{f:venn}}
\end{figure}

What can we learn about the relationship of the ontologies of common sense, physics and mathematics from the examples we analysed? This becomes clear when we use the subdividion of the ontology of physics into system entities, magnitudes and models. In classical physics, which is tightly linked to common sense and in which we understand what we are observing using common-sense thinking, all system entities are real-world entities understandable by common sense. Thus they are either universals in the sense of BFO (see figure \ref{f:venn}, top of left panel) or they are instances of such universals. 

The \x{magnitudes} of classical physics, however, have a dual character. As universals, they are again parts of the coverage domain of BFO (middle of left panel). When viewed from the mathematical point of view, on the other hand, in other words viewed as parts of mathematical models, they are not. Therefore $\textrm{BFO} \cap \textrm{PhysO} = \textrm{PhysO} \setminus \textrm{MathO}$. The models have no BFO part because they do not have instances in reality (bottom of left panel).

To see how this works consider the mass used in the forced harmonic oscillator model with dampening (a realistic model). This is the mass of BFO, a quality of a material entity. 
The magnitude mass used to measure the weight of the entity is this same BFO entity. But the magnitude viewed mathematically is something quite different.
Of course, the harmonic oscillator equation cannot be viewed as we can view the real system with a common sense stance, because it exists only as a mathematical entity, and mathematical entities, as we saw, do not have instances in the real world.

With modern physics (since 1905), the relationship between common sense, physics and mathematics looks very different (see right hand panel of figure \ref{f:venn}). \x{Modern physics has no entities in the sense of BFO,} because it relies on mathematics to define system elements, magnitudes and models. The latter are one and all mathematical entities, which have nothing corresponding to them in common-sense reality, the tuple $\langle \mathrm{mathematical~entity}, \mathrm{measurement} \rangle$ is not discovered but rather declared.

For example, an electron is a BFO:material entity which falls within the coverage domain of classical physics. But in quantum field theory (QFT), an electron is a probabilistic field quant and not a BFO entity of any sort. Because all system elements are made of matter, which is described in  QFT using mathematical entities (non-universals), there are no QFT system elements that are BFO entities. 
All magnitudes in modern physics are grounded either in QFT or in the general theory of relativity (for phenomena which QFT cannot model). Yet all their magnitudes are mathematical entities which do not correspond to any natural universals, but are rather only a matter of entity-measurement tuples which we \x{declare} in the way described in section \ref{magnitudes}. Only in the model branch of the physics ontology is the relationship to mathematics and BFO the same in classical as in modern physics: there are no BFO entities in either case.

\section{Discussion}

The approach presented here differs in several ways from the current attempts to formulate an ontology of physics.
The epistemological foundation of our view is that modern, mathematical physics as a science mediates between our perceptual experience of the world and mathematical \x{a priori} entities which are mind-dependent and do not exist outside the collective consisting of the minds of mathematicians.
It is crucial to understand that such entities exist and to grasp their fundamental role as objects of human thinking as well as the ways we think of them and how we relate them to each other ontologically: not via Aristotelian genus-species-subsumption hierarchies, but using set-theoretical relations to describe their taxonomic relations (and otherwise using the full arsenal of mathematical relations).
Because they essentially use the abstract entities of mathematics, the systems modelled by physics are abstractions of reality in mathematical form. When a physicist creates a model, he gives a description of a system which uses simplified system properties in order to enable the construction of a mathematical model. Once the mathematical model is available, purely mathematical reasoning processes such as variable substitution or term approximation can be used to refine it.

In classical physics the model then has two roles. As a mathematical equation, it is an essential structure which the mathematician uses to manipulate the equation algebraically. But at the same time, it relates real-world magnitudes to each other inside systems. The fact that variables used in mathematical equations are both mathematical entities and represent real magnitudes enables classical physics to mediate between reality and mathematics. Modern theoretical physics, in contrast, cannot achieve this, as there are no available universals. Applied physics can be used to engineer technical systems that can be used to demonstrate the ability of the models to explain and predict aspects of nature. 
They thereby combine classical and modern physics. But even here the system elements from modern physics are not universals either, as we have seen. When engineers use modern physics for engineering, for example when building a laser, they combine instances of universals together to create technical devices using their knowledge of mathematical entities.

Our approach to physics ontology with the division of physics entities into systems, magnitudes and models reflects our understanding of the physicist's knowledge generation process. 
In classical physics, systems are selections from reality which are chosen and delimited as objects of scientific inquiry. Here, the magnitudes are universals. They mediate between the reality of the system and the idealised nature of the model.
We saw that the \x{harmonic oscillator} model illustrates in which way we abstract from reality to create idealised models in physics, but the model can be refined to an extent that it can model reality very closely, for example, to yield the forced oscillator with dampening that models a real electric circuit.
Because the model uses Euclidean space ($\mathbb{R}^3$) as phase space, we can easily imagine the relationship of the model to the sensory reality we perceive in our common-sense view of the world as an environment extending in three spatial and one temporal dimension.
The magnitudes we use to link the model to reality can not only be imagined quite well, they can also be experienced: we can feel our own mass and the acceleration and force acting on our bodies, and we can feel the retraction of a spring and also the passing of time as our heart beats and as we breathe.

With the second system, the \x{entangled photons}, we have again a \x{system that really exists}. But the \x{model} is to a much greater degree a creature of the mind, and though it models our measurements, it does not represent the real system. As we have shown, though the model defined by equation (\ref{e:sup}) has a physical state to which it relates, we do not understand it; in this special case, the model \x{merely expresses the artificial information distribution (what we call `photon entanglement) that we have created using parametric down-conversion with a machine}.

More generally speaking, in modern physics, models merely mediate between the mostly indirect measurement of system entities on one side and the highly abstract essential structures of mathematics on the other. The real part that remains are the measurements to which we have attributed mathematical magnitudes. But we have no universals.

The mathematical knowledge stack that is needed to conceive and understand such models is quite deep and broad, and talented students usually need three to four years to acquire it. The model, as we have seen, can be understood only through the view of the mathematical entities it is made of, and even the one magnitude that it describes -- the spin -- cannot be imagined in the way that we think of magnitudes in classical physics using our natural, common-sense attitude. We cannot think that the particle turns on its own axis, though sometimes the spin is shown like this in textbooks as a pseudo-illustration. Rather, spin is really only a property that we can measure with complicated machines and which we model using quantum projectors (see \ref{proj}). We cannot imagine this property: we can only think of it as a $\mathcal{H}$-space projector.

Therefore, post-classical physics cannot be ontologically represented using common-sense-based ontologies, but requires an ontology of mathematics which gives us a possibility of thinking about the ideal entities of it postulates.
 It is nevertheless a miracle of the human mind that we can use these entities to model and engineer useful machines such as the MRI, the laser, or quantum sensors.


\printbibliography

\end{document}